\documentclass[aps,prl,twocolumn,nofootinbib,a4paper,amsfonts,amssymb,amsmath,floatfix,superscriptaddress]{revtex4-2}
\pdfoutput=1
\usepackage[utf8]{inputenc}
\usepackage[english]{babel}
\usepackage[colorlinks=true,allcolors=magenta,breaklinks=true]{hyperref}
\usepackage{tikz}
\usepackage{graphicx}
\usepackage{dsfont}
\usepackage{soul}
\usepackage{microtype}
\usepackage{mathtools}
\usepackage{amsthm}
\usepackage{microtype}

\begin{document}
\title{The Axiom of Choice and the No-Signaling Principle}

\author{\"Amin Baumeler}
\thanks{All authors contributed equally and are listed alphabetically.}
	\affiliation{Institute for Quantum Optics and Quantum Information (IQOQI), Austrian Academy of Sciences, Boltzmanngasse 3, A-1090 Vienna, Austria}
        \affiliation{University of Vienna, Faculty of Physics \& Vienna Doctoral School in Physics \& Vienna Center for Quantum
Science and Technology, Boltzmanngasse 5, 1090 Vienna, Austria}
        \affiliation{Facolt\`a indipendente di Gandria, 6978 Gandria, Switzerland}
\author{Borivoje Daki\'c}
\thanks{All authors contributed equally and are listed alphabetically.}
		\affiliation{Institute for Quantum Optics and Quantum Information (IQOQI), Austrian Academy of Sciences, Boltzmanngasse 3, A-1090 Vienna, Austria}
		\affiliation{University of Vienna, Faculty of Physics \& Vienna Doctoral School in Physics \& Vienna Center for Quantum
Science and Technology, Boltzmanngasse 5, 1090 Vienna, Austria}
\author{Flavio Del Santo}
\thanks{All authors contributed equally and are listed alphabetically.}
	\affiliation{Institute for Quantum Optics and Quantum Information (IQOQI), Austrian Academy of Sciences, Boltzmanngasse 3, A-1090 Vienna, Austria}
\affiliation{University of Vienna, Faculty of Physics \& Vienna Doctoral School in Physics \& Vienna Center for Quantum
Science and Technology, Boltzmanngasse 5, 1090 Vienna, Austria}

\begin{abstract}
	We show that the \emph{axiom of choice,} a basic yet controversial postulate of set theory, leads to revise the standard understanding of one of the pillars of our best physical theories, namely the \emph{no-signaling principle.}
	While it is well known that probabilistic no-signaling resources (such as quantum non-locality) are stronger than deterministic ones, we show---by invoking the axiom of choice---the opposite: 	Functional (deterministic) no-signaling resources can be stronger than probabilistic ones (e.g. stronger than quantum entanglement or non-locality in general).
	To prove this, we devise a Bell-like game that shows a systematic advantage of functional no-signaling with respect to any probabilistic no-signaling resource.
\end{abstract}

\maketitle
\textit{Introduction.--}As Eugene P.~Wigner famously put it, {\em ``mathematics plays an unreasonably important role in physics''\/}~\cite{wigner1960unreasonable}.
Indeed, the formalization of physical theories is deeply rooted into a plethora of mathematical concepts, some of which lie beyond the grasp of human experience and intuition.
Even within a branch of mathematics that arises naturally like set theory, seemingly self-evident axioms may lead to far-fetched consequences.
It is the aim of this letter to show that examining in detail one of the basic postulates of set theory, the \emph{axiom of choice}~(AC), forces one to rethink  the standard understanding of \emph{no-signaling}~(NS) in physical theories, one of the most fundamental constraints closely related to causality \cite{bell2004speakable}.

The AC states that given any collection of mutually disjoint nonempty sets, it is possible to pick exactly one element \emph{(the canonical representative)} from each member of the collection, and, in turn, to collect them into a~set~\emph{(the choice set).}
While this is straightforward for finite sets (which are always well-ordered), dealing with infinite sets leads to powerful yet controversial results, such as the notorious Banach-Tarski paradox~\cite{banach1924decomposition}.
This makes the AC {\em ``the most discussed axiom of mathematics, second only to Euclid’s axiom of parallels'\/}'~\cite{fraenkel1973foundations}.
In fact, while the adoption of the AC allows proving essential results of mathematics~\cite{howard1998consequences, ash1975consequence} and their applications to physical theories (such as the existence of a basis for every vector space), controversies about the necessity of introducing the AC have been around since Ernst Zermelo~\cite{zermelo1904} introduced it in 1904.
But it was only quite recently, exactly one century after its formalization, that novel counter-intuitive consequences of the AC have been explored, mostly in the context of a class of logic puzzles known as ``infinite hat problems''~\cite{hardin2016mathematics}.\footnote{These results were apparently put forward by two graduate students, Yuval Gabay and Michael O’Connor, in 2004 and further developed most notably by Christopher S.~Hardin~\cite{hardin2008introduction, hardin2008peculiar}.}

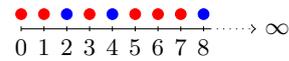
\begin{figure}
	\centering
	\begin{tikzpicture}
		\def\lh{.1em}
		\def\radius{2pt}
		\def\delta{.6em}
		\def\distscale{.3}
		\draw[-] (0,0) -- ++(8.3*\distscale,0);
		\draw[dotted,->] (8.3*\distscale,0) -- ++(2*\distscale,0) node[right] {$\infty$};
		\newcommand{\player}[2]{
			\draw (#1*\distscale,\lh) -- ++(0,-2*\lh) node[below] {$#1$};
			\filldraw[#2] (#1*\distscale,\delta) circle (\radius);
		}
		\player{0}{red}
		\player{1}{red}
		\player{2}{blue}
		\player{3}{red}
		\player{4}{blue}
		\player{5}{red}
		\player{6}{red}
		\player{7}{red}
		\player{8}{blue}
	\end{tikzpicture}
	\newcommand{\hatblue}[0]{\protect\tikz{\protect\filldraw[blue] (0,0) circle (2pt);}}
	\newcommand{\hatred}[0]{\protect\tikz{\protect\filldraw[red] (0,0) circle (2pt);}}
	\caption{The ``infinite hat game:'' Players are positioned on a line from~$0$ to~$\infty$, each player is assigned a hat with a random color (red or blue), and must guess the own hat color by only observing the hat colors of all players in front. For instance, in the current assignment player 3 sees~(\hatblue,\hatred,\hatred,\hatred,\hatblue,\dots).}
	\label{fig:game}
\end{figure}

The standard instance of an ``infinite hat problem''~\cite{hardin2016mathematics} is a theoretical game in which infinitely albeit countably many players are placed on a~ray, one after the other, facing the same direction towards infinity.
For each player a~hat with two possible colors (say red and blue) is picked at random and placed on her or his head.
Then, the players are challenged to guess the color of their own hat by only seeing the colors of the hats of all the (infinitely many) players in front of them, but not the own nor the hat colors of the players behind.
Moreover, the players are not allowed to communicate, but they might agree on a guessing strategy before the game starts (see Figure~\ref{fig:game}).
Since the hat colors are independent and uniformly distributed, the hats of all players excluding a player~$k$ do not contain any information about the hat color of the~$k$th player;
any strategy seems to be equally weak and gives half probability to each player for a correct guess. 
However, and in stark contrast to this naive reading, the AC allows for strategies where {\em all but finitely many players guess correctly.}
The strategy that invokes the AC is the following.
First, we represent the two different colors, red and blue, with bits, zero and one.
The {\em actual\/} assignment of hat colors to the players is thus an infinitely long bit string~$\vec x=(x_1,x_2,\dots)$, with $x_k\in \{0,1\}, \ \forall k$.
Now, define the binary relation~$\sim$ for infinitely long bit strings as
\begin{align}
	\vec y \sim \vec z : \Longleftrightarrow \exists t<\infty, \forall i>t: y_i=z_i
	\label{eq:equivalencerelation}
	\,.
\end{align}
Because this relation is an equivalence relation, the AC allows the players to select a representative from each equivalence class.
To win this game, each player~$k$ translates the observations into an infinitely long bit string~$\vec a^{(k)}=(x_{k+1},x_{k+2},\dots)$, prepends~$k$ zeroes to obtain~$\vec a'^{(k)}=(0,\dots,0,x_{k+1},x_{k+2},\dots)$, invokes the AC to obtain the representative~$\vec r$ of the equivalence class~$\vec a'^{(k)}$~(and therefore~$\vec x$) belongs to, and guesses according to the bit~$r_k$.
By the definition of the equivalence relation, at most finitely many players may output an incorrect guess;
there is a finite~$t$ after which the bits~$x_i$ and~$r_i$ are identical, i.e., $x_i=r_i, \ \forall i> t$.
This unintuitive result has puzzled mathematicians, but so far its implications seem to be discussed only within mathematical logic.
Although it is clear that a scenario like the one described above is at best implausible to find real-world realizations, we deem it important to discuss its consequences in the framework of the fundamental principles of physics.
After all, the whole of the theoretical apparatus of physics relies heavily on the use of set theory, thus on the AC.

\begin{figure}
	\centering
	\begin{tikzpicture}
		\def\boxw{2em}
		\def\boxh{\boxw}
		\def\sep{4em}
		\def\llen{.8em}
		\draw[thick,fill=black] (-\sep,-\boxh/2) rectangle ++(-\boxw,\boxh);
		\draw[thick,fill=black] (+\sep,-\boxh/2) rectangle ++(+\boxw,\boxh);
		\draw[->] (-\sep-\boxw/2,\boxh/2) -- ++(0,\llen) node[above] (a) {$a$};
		\draw[->] (\sep+\boxw/2,\boxh/2) -- ++(0,\llen) node[above] (b) {$b$};
		\draw[<-] (-\sep-\boxw/2,-\boxh/2) -- ++(0,-\llen) node[below] (x) {$x$};
		\draw[<-] (\sep+\boxw/2,-\boxh/2) -- ++(0,-\llen) node[below] (y) {$y$};
		\draw[dashed] (0,0) -- ++(45+0*90:3*\boxw);
		\draw[dashed] (0,0) -- ++(45+1*90:3*\boxw);
		\draw[dashed] (0,0) -- ++(45+2*90:3*\boxw);
		\draw[dashed] (0,0) -- ++(45+3*90:3*\boxw);
	\end{tikzpicture}
	\caption{As dictated by causality (the dashed lines represent light cones), the behavior~\mbox{$P(a,b\mid x,y)$} of two space-like separated black boxes satisfies the no-signaling constraint: Changing the input~$x$~($y$) on one side has no effect on~$b$~($a$) on the other side.}
	\label{fig:ns}
\end{figure}
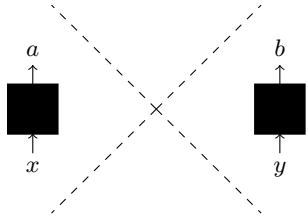

\textit{Probabilistic no-signaling.--}We recast the above ``infinite hat problem''  using a device-independent approach, as customarily done in modern quantum information~(see, e.g.,~\cite{brunner2014bell} and references therein).
This framework aims at modeling physical problems in terms of ``black boxes''~(i.e.,~abstract devises without specifications of~ the involved physical systems, mechanisms, or degrees of freedom) with inputs and outputs.
Different classes of theories are then singled out by characterizing the joint probability distribution of the outputs given the inputs, under certain constraints.
A~prime example thereof is the \emph{no-signaling} constraint~(see Figure~\ref{fig:ns}).
For two parties, $A$ and $B$, each provided with a black box, who both each receive an input, $x$ and $y$, and whose box returns an output, $a$ and $b$, respectively, the NS constraint reads
\begin{align}
	\begin{split}
		\forall x: \sum_a P(a,b \mid x,y)&= P(b\mid y)\,,\\
		\forall y: \sum_b P(a,b \mid x,y)&= P(a\mid x)\,.
	\end{split}
	\label{nsprob}
\end{align}
This intuitively encapsulates the concept of causality, in so far as changing one party's input  does not change the probability of the outputs observed on the other side.
In fact, when embedded in space-time, the no-signaling constraint must hold between any two space-like separated  input-output pair of events, as otherwise the players could communicate faster-than-light.
While this definition seems {\it prima facie\/} unproblematic, we would like to point out that it relies on the mathematical construct of probability theory, which carries along a baggage of mathematical assumptions that we usually take for granted.
In particular, a probability (as defined through the Kolmogorov axioms \cite{kolmogorov2018foundations}) is a measure function $P$ defined on a~probability space~$(\Omega, \Sigma, P)$, where~$\Omega$ is a sample space of events and~$\Sigma$ is a~$\sigma$-algebra.
So, for expressions~\eqref{nsprob} to make sense, $x$, $y$, $a$, and $b$ should be values of well-defined random variables.
In this letter, we ask:
Is it always possible to define NS based on probability distributions of outputs given inputs? 

We will reply in the negative due to the fact that the AC  entails the existence of non-measurable sets, and thus of in general not well-defined random variables.
To see this, consider the following simple example.
Let~\mbox{$f:\mathbb R\cap [0,1]\rightarrow\{0,1\}$} be an arbitrary function.
It follows that $f^{-1}(0) \in [0,1]$ and $f^{-1}(1) \in [0,1]$, and since they exhaust all the possible cases:~$f^{-1}(0) \cup f^{-1}(1) = [0,1]$.
But since the AC guarantees the existence of functions that do not map measurable sets to measurable sets,~$f^{-1}(0)$ and~$f^{-1}(1)$ are in general not measurable.
Therefore, if~$x\in\mathbb R\cap [0,1]$ is a random variable,~$f(x)$ is not necessarily a random variable.
But then one faces the following tension.
On the one hand, the definition of~NS relies on the concept of probability, that in turn is rooted in (measurable) set theory that tacitly assumes the AC.
On the other hand, as we have just seen, the AC entails the existence of non-measurable sets and thus does not guarantee that we can use random variables with well-defined probability distributions.
So, how can one define NS in the presence of the AC?

\textit{Functional no-signaling.--}To address this issue, one can go beyond the concept of probabilistic NS, and define it using exclusively deterministic functions.
In this case, one does not introduce probabilities at all, but rather considers the outputs as functions of the inputs, i.e.,~for~$N$ parties,~\mbox{$\vec a=\left(f_1(\vec x), f_2(\vec x), \dots, f_N(\vec x) \right)$}, where~$\vec x$ is the vectors of inputs~$\vec{x}=(x_1, x_2, \dots, x_N)$ and~$\vec a$ the vector of outputs $(a_1, a_2, \dots, a_N)$, respectively.
At the functional level, the NS condition---which we call {\em functional no-signaling\/} (FNS)---is defined as~\cite{pearl2009}
\begin{align}
	f_k(y_1, \dots, x_k, \dots, y_N)= f_k(z_1, \dots, x_k, \dots, z_N)
	\,,
	\label{functionalns}
\end{align}
for all possible inputs $x_k$, $\vec y$, and $\vec z$.
Note that FNS is the natural way to define no-signaling in a deterministic world, one in which probabilities do not necessarily need to be introduced:
When probability distributions are well-defined, resources satisfying FNS~(Eq.~\eqref{functionalns}) corresponds to the deterministic extremal points of the probabilistic no-signaling resources~(Eq.~\eqref{nsprob}), where all probabilities are either zero or one.
Therefore, when probabilities are well-defined, the set of FNS resources is a proper subset of the set of probabilistic NS resources.
It is interesting to point out that the condition of \emph{locality}---which in the probabilistic case is a stricter condition than NS~\mbox{\cite{khalfin1985,popescu1994,brunner2014bell}}---at the functional level reads $f_k(\vec x)=F_k(x_k)$, and it is in this case equivalent to the FNS condition.\footnote{That functional locality implies FNS is straightforward.
	To see the converse, pick some arbitrary~$\vec y$ and define the function $F_k(x_k) :=  f_k(y_1, \dots, x_k, \dots, y_N)$.
}

Despite this, in what follows we introduce an infinite hat game to show that probabilistic NS and FNS are in general {\em inequivalent\/} resources, and that---contrarily to intuition---FNS is a stronger resource than probabilistic~NS.

\textit{A game of chance.--}Inspired by the hat game previously explained, we formulate a game of chance using the familiar language of modern quantum information, i.e.,~in a similar fashion as quantum communication and non-local games~(see,~e.g.,~\cite{winter2010usefulness, almeida2010guess, briet2013multipartite, del2018two}).
Consider again infinitely many players~$\{1,2,\dots\}$,
each of whom receives an input $x_k\in \mathbb{R} \cap [0,1]$ from a referee.
The referee starts by generating the root~$x_{0}\in\mathbb R\cap [0,1]$ from the unit real interval picked at random with a uniform distribution.
Then, the referee generates correlated inputs through the one-dimensional \emph{baker's map}~$\mathcal B$~(see,~e.g.,~\cite{gisin2019indeterminism}) as follows:
\begin{align}
	x_{k+1}
	:=
	\mathcal B(x_{k}) =
	\begin{cases}
		2x_k &\text{if} \ x_k < \frac{1}{2}\,,\\
		2x_k-1 &\text{otherwise.}
	\end{cases}
	\label{baker}
\end{align}
This transformation maps the real number expressed in its binary expansion~$x_k=0.b_1b_2b_3\dots$ to the real number~\mbox{$x_{k+1}=0.b_2b_3\dots$}, effectively erasing the most significant bit.
Each player is then required to return a binary output~\mbox{$a_k\in \{0,1\}$} to the referee such that it is equal to the most significant bit of the input to the previous player (equivalently, the~$k$th player must guess the~$k$th bit in the binary expansion of the root~$x_0$), i.e.,~the output of the~$k$th player must satisfy the following predicate
\begin{align}
	a_k = 2x_{k-1}-x_{k}
	\,.
	\label{eq:predicate}
\end{align}
Notice that this game is nothing else but a variant of the well-know ``guess your neighbor's input'' game~\cite{almeida2010guess}.
Our aim is to find the strategy that maximizes the amount of successful players, given only no-signaling resources.

Let us start our analysis within the probabilistic framework. Let us denote by $\vec{x}$ and $\vec a$ the (infinite-dimensional) vectors of all the inputs and all the outputs, respectively.
The probability that the $k$th player provides the requested output is
\begin{align}
	P_{\text{win}}^{(k)} &= \int_0^1 dx_0 \sum_{\vec a \mid a_k = 2x_{k-1}-x_{k}} P(\vec a | \vec x)\\
	&= \int_0^1 dx_0 \sum_{ a_k = 2x_{k-1}-x_{k}} \sum_{\vec a_{\setminus k}} P(\vec a | \vec x)
	\,,
	\label{eq:pwin}
\end{align}
where we have used the fact that $x_0$ is uniformly distributed and that all the other inputs are functions of~$x_0$;~$P(\vec a | \vec x)$ is the overall probability of finding outputs $\vec a$ given the inputs $\vec{x}$.
In the second line we have split the sum into one on the index $k$ and another on all the other outputs $\vec a_{\setminus k}=(a_1,\dots,a_{k-1},a_{k+1},\dots)$.

We now impose the no-signaling constraint.
This, generalizing Eq.~\eqref{nsprob}, translates into the standard condition on the marginal probability distribution as follows:
\begin{align}
	\forall \vec x: \sum_{\vec a_{\setminus k}} P(\vec a | \vec x)= P(a_k|x_k) \,.
	\label{eq:NS}
\end{align}
With this constraint, Eq.~\eqref{eq:pwin} reads:
\begin{align}
	&P_{\text{win}}^{(k)}
	= \int_0^1 dx_0 \sum_{ a_k = 2x_{k-1}-x_{k}} P(a_k | x_k)
	\,.
	\label{eq:pwin2}
\end{align}
The domain of integration can now be split into two parts, in which $a_k$ takes the value zero or one, respectively, and moreover we express~$P_{\text{win}}^{(k)}$ as a function of $x_{k-1}$:
\begin{align}
	P_{\text{win}}^{(k)}
	= &\int_0^{1/2} dx_{k-1} \mu(x_{k-1}) P(0\mid \mathcal B(x_{k-1})) +\\
	&\int_{1/2}^1 dx_{k-1} \mu(x_{k-1}) P(1\mid \mathcal B(x_{k-1}))
	\,,
	\label{eq:pwin2bis}
\end{align}
where~$\mu(x_{k-1})$ is the probability distribution of the input~$x_{k-1}$.
Since the uniform distribution is invariant under the application of~$\mathcal B$, $\mu(x_{k-1})=1$,\footnote{To show this, consider a real number $y \in \mathbb{R} \cap [0,1] = \mathcal B(x)$, where the distribution of $x$ is uniform, i.e.,~$\lambda(x)=1$.
The cumulative distribution function reads $P(y\leq q)=\int_0^q \mu(y)dy=\int_0^{x_1} \lambda(x)dx+\int_{1/2}^{x_2} \lambda(x)dx$, for some $x_1$, $x_2$ and $q=\mathcal B(x_1)=\mathcal B(x_2)$.
But since $\lambda(x)=1$, $\int_0^q \mu(y)dy=q$; taking the derivative with respect to $q$ on both sides, one finds $\mu(y)=1$, which completes the proof.}
and the winning probability reads:
\begin{align}
	&P_{\text{win}}^{(k)}
	= \int_0^{1/2} dx_{k-1} P(0\mid 2 x_{k-1}) +\\
	&\hspace{3em}\int_{1/2}^1 dx_{k-1} P(1\mid 2 x_{k-1}-1)\\
	&= \int_0^{1/2} d x_{k-1}
	\left(
	P(0\mid 2 x_{k-1})
	+
	P(1\mid 2 x_{k-1})
	\right)\\
	&=\frac 1 2
	\,.
	\label{eq:pwin3}
\end{align}
Hence, resorting exclusively to probabilistic no-signaling strategies, each player outputs the correct value with half probability.

In order to compute the optimal overall strategy, i.e.,~the one that maximizes the amount of winning players, let us define for each player a ``success'' random variable~$s_k$, where $s_k=1$ means that the $k$th player wins the game and $s_k=-1$ that it fails.
The sequence of random variables $S_N=\sum_{k=1}^n s_k$ gives a (finite) estimation of the amount of players that win the game.
Given Eq.~\eqref{eq:pwin3}, it is straightforward to check that $S_n$ is a martingale, i.e.,~$|S_{n+1}-S_n|<\infty$, and the expectation value satisfies
\begin{align}
	E[S_{n+1} \mid S_1,S_2,\dots,S_n]
	&=E\left[\sum_{k=1}^{n+1} s_k\right]\\
	&=S_n+E[s_{n+1}]=S_n
	\,.
\end{align}
Therefore, Azuma's theorem~\cite{azuma1967weighted} applies, yielding
\begin{align}
	 \Pr\left[S_n \geq \varepsilon\right]\leq 2 e^{-\varepsilon^2/2n}
	 \,,
\end{align}
for every real number $\varepsilon>0$.
This means that, in the limit of~$n$ approaching infinity, the probability of deviation from the average---corresponding to a probability of success for each player of 1/2---exponentially vanishes.
Hence, imposing probabilistic NS, the ratio of players that win the game over the total converges to 1/2 almost surely.

\textit{Functional NS violates the boundary of any probabilistic NS resources.--}Let us now turn to a strategy that exploits the axiom of choice, in the same fashion of the hat problem previously recalled.
Formally, the AC states that given a collection $C$ of sets $X$, such that $\emptyset \notin C$, there exists a \emph{choice function}~$f$ on $C$, defined by the property~$f(X)\in X$.
In our case, let $C$ be the collection of the equivalence classes of infinitely long bit strings under the relation~$\sim$~(see Eq.~\eqref{eq:equivalencerelation}).
We have already noticed that, given the form of the correlated inputs through Eq.~\eqref{baker}, the inputs~$2^k x_k\in\mathbb R$ belong all to the same equivalence class~$X$.
The AC thus ensures the existence of a function with the following property
\begin{align}
	\forall x_k \in X:
	f(2^k x_k)=r \,,
	\label{eq:choicef}
\end{align}
where~$r$ is a specific element of $X$ called the canonical representative of $X$.

\begin{figure}
	\centering
	\includegraphics[width=8cm]{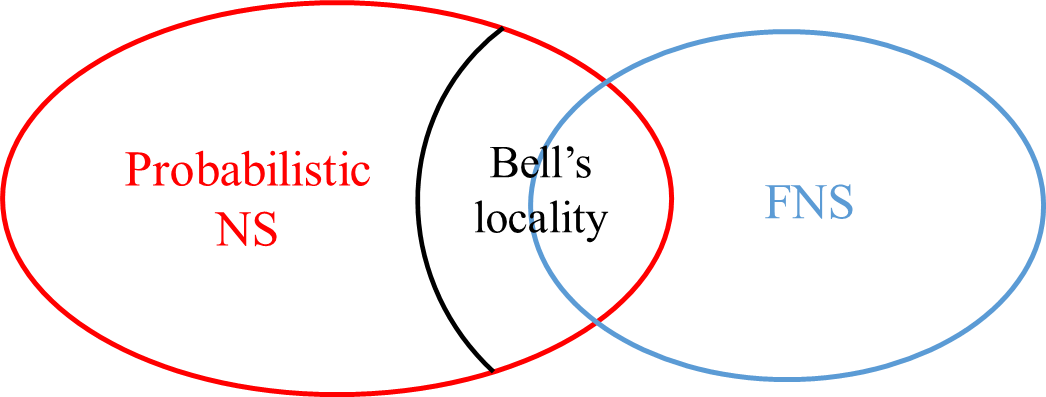}
	\caption{Suggestive representation of the no-signaling resources. The overlap of probabilistic NS and FNS are the deterministic resources over random variables.
		Bell's locality encapsulates Bell-local resources that are not deterministic, {\it e.g.,\/}~shared coin flips. Quantum-theoretic and general violations of Bell inequalities represent resources that are probabilistic NS yet not FNS. Here, we provide an opposite distinction:
	The game proposed shows the existence of FNS resources that are not probabilistic NS.}
	\label{fig:nssets}
\end{figure}

Now, let us consider that each player has access to a~local black box, an \emph{oracle,} that implements the choice function~\eqref{eq:choicef}.
In this way, by means of only local operations and without any communication between the players, each of them may output
\begin{align}
	a_k = \lfloor{2 \mathcal B^{k-1}(f(2^k x_k))}\rfloor
	\,.
	\label{eq:outcome}
\end{align}
With this strategy, all players $k> t$ for some finite threshold~$t<\infty$ return the correct answer according to the predicate~\eqref{eq:predicate} with certainty.
Hence, all but~$t$ players (i.e.,~an arbitrarily large but finite number of them) win the game without violating FNS (as straightforwardly shown by Eq.~\eqref{eq:choicef}).
Remarkably, this violates the boundary found by imposing any possible probabilistic NS strategy,
and proves that FNS is in general a {\em stronger\/} resource than probabilistic NS (see Figure~\ref{fig:nssets}). 

\textit{Outlook.--}We have investigated the consequences of the fundamental postulate of set theory known as the axiom of choice for no-signaling in physical theories.
We showed that the AC forces us to redefine NS without resorting to probabilities but only in terms of deterministic functions.
By introducing a game of chance, we showed that the standard probabilistic NS and the functional NS are not equivalent resources and that, surprisingly, the latter are in general stronger resources than the former.
Our result goes in the same direction of a resolution of Bell's non-locality proposed by Pitowsky, who concluded that {\em ``the violation of Bell's inequality reflects a mathematical truth, namely, that certain density conditions are incompatible with the existing theory of probability''\/}~\cite{pitowsky1982resolution}.
More at the philosophical level, resorting to functions to define no-signaling (FNS) could be seen as the natural description in a fully deterministic world, where probabilities are not fundamental and therefore disposable.
On the other hand, in an indeterministic world in which uncertainty is quantified by probability, deterministic behaviors are retrieved as extremal cases from the more general probabilistic ones.
We have shown that these two concepts of deterministic no-signal are not equivalent resources.
Moreover, it ought to be remarked that in the standard formulation of quantum theory no deviation from the winning probability of~$1/2$ is possible: The probabilistic NS strategies contain the quantum strategies.
We leave open whether a game exists that displays a separation between Bell-local and quantum, as well as quantum and FNS resources, and---towards generalizing quantum theory---how and in what sense quantum theory and especially Born's rule could be extended to incorporate the AC.
The latter question might yield novel insights into the indeterministic nature of quantum theory. 


One could object that our work makes use of the concept of infinity, and that this could be problematic when discussing physics~\cite{ellis2018physics, gisin2019indeterminism, del2019physics, del2021relativity}. Note, however, that one could equally argue that operationalizing the concept of mathematical probability---which is ubiquitous in modern physics---also requires infinite repetitions.

More interpretational questions remain open.
What are the implications of our result for probability theory and its interpretations?
The power of the AC seems to uncover a new form of uncertainty:
While clearly the hat colors of all players but the~$k$th contain no information of the~$k$th, for all but finitely many players it is possible to {\em extract\/} that information from the infinite tail of the color sequence.
This is in tension with the Bayesian interpretation of probability \cite{de2017theory} where any uncertainty is modeled probabilistically.

According to our most successful theories, no-signaling is perhaps the most fundamental limit of nature, so studying the consequences of its (perhaps overlooked) mathematical properties is of prime importance for our understanding of physics and its limits.

\acknowledgments
\textit{Acknowledgments.--}We thank Charles-Alexandre B\'edard, Xavier Coiteux-Roy,  Kyrylo Simonov, and Stefan Wolf for insightful discussions. B.D. thanks Milovan {\v S}uvakov for introducing him to the original hat puzzle.
\"A.B.~acknowledges support from the Austrian Science Fund (FWF) through ZK3 (Zukunftskolleg) and through BeyondC-F7103.
B.D.~also acknowledges support from FWF through BeyondC-F7112. F.D.S.~acknowledges support from FWF through the project “Black-box quantum information under spacetime symmetries”, OFWF033730.

\bibliography{refs}
\end{document}